\documentstyle[aps,prl]{revtex}

\begin{document}

\title{Comment on ``London Theory for Superconducting Phase Transitions in External Magnetic Fields: Application to $\text{UPt}_{3}$''}

\author{ V.P. Mineev and T. Champel}
\address{ Commissariat \`{a}
l'Energie Atomique, DSM/DRFMC/SPSMS\\
17 rue des Martyrs, 38054 Grenoble, France}

\date{\today}

\maketitle

The theory of the equilibrium flux lattice in $UPt_{3}•$ developed in Ref. \cite{1}
is valid for the superconducting A-phase near the phase
transition to the normal state, where the symmetry of the superconducting
state is mostly dictated by one of the two components of the order parameter $\bbox{\eta}=\left(\eta_{1},\eta_{2}\right)$.  An attempt to spread the analysis
\cite{1} towards the phase transition between the superconducting A and B
states has been undertaken in the recent Letter \cite{2}.  In our
opinion, the approach and the conclusions of Ref. \cite{2} are wrong and
here we present why it is so.

The Ginzburg-Landau equations for $H \parallel z$  after taking just for brevity $\beta_{2}•=0$ are \cite{3}
\begin{eqnarray}
\left(\alpha_{0}\tau_{1} +2\beta_{1} \left(\left|\eta_{1}\right|^{2}
+\left|\eta_{2}\right|^{2}\right)
+\left(K_{123}D_{x}^{2}+K_{1}D_{y}^{2}\right)\right)\eta_{1} =\nonumber\\
-\left(K_{2}D_{x}D_{y}+K_{3}D_{y}D_{x}
\right)
\eta_{2},\\
\left(\alpha_{0}\tau_{2}+2\beta_{1} \left(\left|\eta_{1}\right|^{2}
+\left|\eta_{2}\right|^{2}\right)
+\left(K_{123}D_{y}^{2}+K_{1}D_{x}^{2}\right)\right)\eta_{2} =\nonumber \\
-\left(K_{3}D_{x}D_{y}+K_{2}D_{y}D_{x}
\right)
\eta_{1},
\end{eqnarray}
where
$\tau_{1}=(T-T_{1})$,  $\tau_{2}=(T-T_{2})$ with the temperatures $T_{1} > T_{2}$.
Near the phase transition to the normal state, that is when $H\to
H_{c2}•=-\Phi_{0}•\alpha_{0}•\tau_{1}•/2\pi\sqrt {K_{1}•K_{123}•}$ the
component $\eta_{2}$  is expressed through the first
component $\eta_{1}$ \cite{1}:
\begin{equation}
    \eta_{2}=-\frac{\left(K_{3}D_{x}D_{y}+K_{2}D_{y}D_{x}
\right)\eta_{1}} {\alpha_{0}\left(T-T_{c2}\right)},
\end{equation}
and $\eta_{1}$ is solution of the nonlinear Eq. (1) at $\eta_{2}•=0$
 (here $T_{c2}=T_{2}-2\beta_{1} \langle \left|\eta_{1}\right|^{2} \rangle/\alpha_{0}$). 
 
It is clear that in the
region of the neighbourhood of the AB transition ($T\to T_{c2}•$), the simple linear but nonlocal relationship (3) between the two order
parameter components is divergent and out of applicability.  
The grow of the amplitude of $\eta_{2}•$ near
the transition between the  A and B states is limited in fact principally by the
nonlinear (and local) terms in the free energy.  The mixing gradient terms (right hand side of Eq (1)-(2))
transform the AB and AC phase transition lines into regions of crossover
between the superconducting states with the same symmetry.  The
crossovers are certainly not characterized by a divergence of the
corresponding coherence length as it has to be in the case of a real second order phase
transition.

The authors of the Letter~\cite{2} disregard the contribution of the nonlinear term $2 \beta_{1}\left|\eta_{2}\right|^{2} \eta_{2}$ in Eq. (2) by considering for $H \sim H_{c1}$ (with $K_{2}=K_{3} \ll K_{1}$) the relationship 
\begin{equation}
\left( \alpha_{0}\left(T-T_{c2}\right) +K_{1}{\bf D^{2}}\right) \eta_{2}=-K_{2}\left(D_{x}D_{y}+D_{y}D_{x}
\right)\eta_{1}.
\end{equation}
The divergence at $T \to T_{c2}$ is avoided here by the presence of the differential operator acting on $\eta_{2}$.
Substituting Eq. (4) back to the gradient part of the free
energy, Agterberg and Dodgson have found an effective nonlocal
(4-th order in gradients) London energy functional for fields and
currents originating from 
$\eta_{1}•$. 
These additional nonlocal terms in the London energy starts to be important
when $T \to T_{c2}$.  
As the results, the authors of Ref. \cite{2} have
derived many conclusions concerning the structures of the Abrikosov
lattice in vicinity of the AB transition.

It is possible to estimate under which conditions the nonlinear term is negligible at $T \sim T_{c2}$. Using Eq. (4), we have
\begin{equation}
\frac{\beta_{1} \langle \left|\eta_{2}\right|^{4} \rangle}
{K_{1}\langle \left| {\bf D}\eta_{2}\right|^{2}\rangle} \sim \left(\frac{K_{2}}{K_{1}}\right)^{2} 
\kappa^{2} \left(\frac{\lambda}{d}\right)^{4},
\end{equation}
where $\kappa \approx 60$ is the Ginzburg-Landau parameter for $UPt_{3}$, $d$ the distance between vortices and $\lambda$ the London penetration depth. As $H \to H_{c1}$, $d \sim \lambda (\ln \kappa)^{-1/2}$ so that the nonlinear term is
found negligible when
\begin{equation}
\frac{K_{2}}{K_{1}} < \left(\kappa \ln \kappa \right)^{-1}.
\end{equation}
On the other hand, the approach of Ref.~\cite{2} does not consider the usual nonlocal terms originating from the 4-th order gradient terms acting on $\eta_{1}$ in the free energy, which is valid if 
\begin{equation}
K_{2}/K_{1} > \kappa^{-1}.
\end{equation}
Taking account of (6) and (7), 
we thus conclude that the nonlinear terms are important in the vicinity of the AB crossover.

\end{document}